\documentstyle{article}
\begin{document}
\title
{Quantum Mechanics in General Relativity\\
and its  Special--Relativistic Limit}\thanks{gr-qc/9807030}
\author {\large E.A.Tagirov \\
 N.N.Bogoliubov Laboratory of Theoretical Physics\\
Joint Institute for Nuclear Research, Dubna, 141980, Russia \\
   e--mail: tagirov@thsun1.jinr.dubna.su }
\maketitle

\begin{abstract}
Having started with the general formulation of the quantum  theory
of the real scalar field  (QFT) in the general Riemannian space--time
$\,V_{1,3}\,$, the  general--covariant quasinonrelativistic quantum
mechanics of a point-like spinless particle in $\,V_{1,3}\,$ is constructed.
To this end, for  any normal geodesic 1+3--foliation of $\,V_{1,3}\,$ ,
a space $\Phi^-$ of asymptotic in $c^{-1}$  solutions of the field equation
is specified, which can be mapped to a space $\Psi$ of solutions of a
Schr\"odinger equation  with an (asymptotically) Hermitean hamiltonian and
the Born probabilistic intepretation of the vectors of $\Psi$. The basic
operators of the momentum and the spatial position of the particle acting in
$\Psi$ generated by the corresponding observables of QFT include relativistic
corrections, and therefore differ generally from those which follow for
the geodesic motion in $\,V_{1,3}\,$ from the canonical postulates of
quantization. In particular, the operators of coordinates do not commute as
well as the  operators of the conjugate momenta, except the cases of
Cartesian coordinates in the Minkowski space--time or of the exact
nonrelativistic limit $(c^{-1} = 0)$. Thus, the shift of quantization
procedure from mechanics to the field theory leads to a more rich
structure of observables in quantum mechanics.
\end{abstract}

\newpage

\newcommand{\Oc}{O\left(c^{-2(N+1)}\right)}
\newcommand{\Sg}{ \Sigma}
\newcommand{\bgt}{ \bigotimes}
\newcommand{\ptl}{ \partial}
\newcommand{\Schr}{ Schr\"odinger representation}
\newcommand{\Schp}{ Schr\"odinger picture}
\newcommand{\Sche}{Schr\"odinger equation\ }
\newcommand{\eu}{$\ E_{1,3}\ $}
\newcommand{\rif}{ V_{1,3} }
\newcommand{\ri}{$\ V_{1,3} \ $ }
\newcommand{\ov}{\overline}
\newcommand{\stc}{\stackrel{def}{=}}
\newcommand{\h}{\hbar}

{\large\bf 1. Introduction}\\

The present paper is devoted to construction of  quantum mechanics
of a particle in the general external gravitational field which is treated
general--relativistically as the metric of {\it the Riemannian space--time}
\ri. Quantum mechanics in \ri has its own domain of application,  at least
speculative, but here I pursue the aim to look at the well--known
but still rather mysterious theory from  the viewpoint of the changed
geometrical background. One may hope that  thus some new knowledge
can be achieved  on the  still rather mysterious  quantum theory. To
my opinion, the results  of the present paper  justify such an expectation
and they concern not only quantum mechanics in \ri, but also  in
{\it the Minkowskian space--time} \eu.

There are two basically different approaches to the problem that has been
set.  The first, more traditional one is quantization of the classical
mechanics. In the simplest case of a neutral and spinless point particle
one should quantize the mechanics of the geodesic motion. The
second, on which the present paper is concentrated, can be characterized as
a restriction to the one--particle configurational subspace
of the  Fock space in the quantum theory of the (linear) field which
corresponds to the particle.
 Simply speaking, I consider a particle as a  spatially localized
configuration of the quantum scalar field (QFT)  and suppose
that creation and annihilation  of particles by the external gravitation
are negligible. Then,
the one--to--one  particle matrix elements of naturally and, in a
sense, uniquely determined in the  operators  of momentum
and spatial coordinate can be represented as  matrix elements of Hermitean
differential operators on a  space $\Psi$ of solutions $\psi (x)$ of
a \Sche   with the  hamiltonian which is Hermitean  with respect
to the inner product induced by  an  $L_2 (\Sg;\,C)$ norm, where $\Sg$  is
{\it a Cauchy  hypersurface} in \ri. This $L_2 (\Sg;\,C)$ norm provides
$\psi (x)$  by the Born probabilistic interpretation  in the configurational
space, and the structure formed by  $\Psi$ and  the Hermitean
operators induced from the QFT is similar to the standard nonrelativistic
quantum mechanics (NRQM), but the velocity of light $c$ is finite in it.
I shall refer  further to this  structure as {\it the quasinonrelativistic
representation} of the  relativistic quantum mechanics. In
the general \ri where the metric depends on time in any system of
reference (see definition in Sec.4) this representation can be constructed
only in the form of asymptotic expansions in $c^{-2}$, which are
valid just for the case  where   creation and annihilation
 of particles can be neglected. In the globally static \ri,
the number of particles does not change, the representation has
a closed form and  is valid formally  for any value of $c^{-2}$
(see Sec.6.).

A question of fundamental importance is:  do
these two approaches, quantization of the classical
mechanics  and the field--theoretical approach,  lead
in any sense to the same quantum mechanics?   In spite
of that in \ri many other questions remain open  in  both the approaches,
the answer to the question posed is definitely negative because
there are distinctions between the resulting structures of quantum
mechanics even in the limiting case of free motion in \eu.

The main distinction is in that the quasinonrelativistic  operators
of momenta and (curvilinear) spatial coordinates   in  $\Psi$,
which follow from the QFT, are different, in general,
from the canonical ones which are  postulated  for the
immediate quantization of mechanics , see Sec.2; exceptions are the  cases
of $c^{-1} = 0 $ for any \ri and  of Cartesian coordinates and momenta
conjugate to them for free motion in \eu.
In particular, with  the indicated exceptions ,  any field--theoretically
defined quasinonrelativistic operators of  coordinates
are noncommutative whereas commutativity of them is  a postulate of
of canonical quantization of mechanics.

 In the case  of free motion in \eu
the representation space $\Psi$ is the space of the negative--frequency
relativistic
wave functions in the Feshbach -- Villars representation [1]. However,
Feshbach and Villars took the canonical expressions for  operators of the
momentum and Cartesian coordinates in fact postulatively whereas I deduce
them from naturally and, in a sense, uniquely defined  QFT--operators.
Besides, remaining on the level of primary quantized theory
Feshbach and Villars considered  the complex scalar field
as describing  electrically charged particles. However, a
complex structure of representation space is a general property
of quantization (this is well explained, for example, in [2])  and
second quantization of a linear field theory consists essentially in
specification of a space of complex  solutions of the field equation
if even  the field were real in the initial classical theory
and the corresponding particles ("quanta" of the field) are not charged.
I consider the neutral scalar field because only chargeless point particles
move in \ri  along   geodesics and
the classical dynamics of electrically charged particles  in \ri
is essentially different  from
the geodesical one even locally as it was shown by Hobbs [3]
who corrected  results of Brehme and DeWitt [4].

Thus, that the change of geometrical background of quantum theory to the
Riemannian one leads to the mentioned unusual conclusions  should not
seem strange if one  recalls   that the canonical quantization
is only a postulate for restoration of a quantum theory from
its classical counterpart
up to $O(\h^2)$. The deformation quantizations, popular now,
and geometries are in fact
attempts to go beyond the limits of this postulate.

Few words on other attempts of field--theoretical approach to  quantum
mechanics in the general external gravitational field. (There is
also a great activity in study of  cases of particular space--time
simmetries, but  I concentrate here on the generally nonsymmetric
space--time specifically  with the hope that it may reveal more distinctly
the role of the symmetry  in the quantum theory.)  A systematic
study  of  the problem on the level of the "primary   quantized"
theory  is done by Gorbatzevich [6] who applied  in  \ri  the operator
method  by Stephani [7]  of transformation  of   the Dirac
equation    to the form of Pauli equation originally done for the case
of the external electromagnetic field in \eu. The Stephani representation
differs essentially from the more known Foldy--Wouthuysen
one in  that it, contrary to the latter, leads to the
$L_2 (E_3;\,C) \oplus L_2 (E_3;\,C) $ norm  of the representation space.
The present paper  emphasizes in particular the necessity  of
$L_2 (\Sg;\,C) $ structure ($\Sg$ is a space--like hypersurface
in \ri, the configurational space
of a particle) for the particle representation of QFT, the fact
seeming obvious but often not recognized.

However, in [6], as well as in [7], the
quantum--mechanical operators of observables are introduced
"manually"  as the canonical
ones. Contrary to this,   in papers  [8--10] of the present author
the  mean values of  quantum--mechanical observables were defined
as natural quadratic functionals of the relativistic
field corresponding to scalar and Dirac particles and
these functionals are expressed as the diagonal matrix elements
of  Hermitean  operators acting  in the general--relativistic
generalizations of the Feshbach -- Villars and
Stephani representation spaces respectively.  In the
present paper this construction for the real scalar field is
consecutively justified  on the basis of the quantum theory
of the real scalar field and its consequences are studied.

The paper is organized as follows. In Sec.2 the postulates of
quantization of a finite--dimensional Hamiltonian system are
recalled and results of formal general--relativistic
application of these postulates to the mechanics of a spinless
particle in \ri are presented according to Sniatycki [11].
However, this approach, being mathematically impeccable,
does not provide with the standard probabilistic  interpretation
of the vectors of representation space which is essentially based
on an 1+3--foliation of \ri on time and space.  Quantization of
the geodesic  motion  in \ri in  the 1+3--formalism is  not
developed yet but the results in [11] suggest the general form of operators
of basic mechanical observables in the 1+3--foliation formalism too.

Further, in Sec.3,   the general set of  Fock representations of
the canonically quantized real scalar field  in the general \ri is considered
and  QFT--operators of basic observables are introduced. The main problem
here is specification of the Fock space, the vectors of which have
a particle interpretation. In Sec.5 its asymptotic solution with respect
$c^{-2}$ is proposed, which is based on the idea on a quantum particle as
a stable field configuration  localized on the normal geodesic
translations $S(x) =const$ of a given  initial
{\it  Cauchy hypersurface $\Sg$}.

In Sec.5 the   matrix elements of the introduced QFT--operators
of spatial position and momentum between the asymptotic  one--particle
states are represented  as matrix elements  of Hermitean (self--adjoint)
differential operators in the space of solutions of a Schr\"odinger equation
in the configurational representation determined by the introduced $\Sg$.
The obtained structure looks as a generally covariant generalization of the
standard NRQM in the Schr\"odinger
representation, but the differential operators of position and momenta
acquire relativistic (asymptotic in $c^{-2}$) corrections of any given
order $N$.

In Sec.6 the  cases of globally static \ri and, in particular, of \eu are
considered. In these cases  a particular normal geodesic congruences
(the  frames of reference) exist in which  the external gravitation
do not change the number of particles. Just in these cases,  the asymptotic
expansions   can be converted
for $\ N\rightarrow \infty\ $ to formally closed exactly relativistic
expressions.
Nonlocality of relation between relativistic and quasinonrelativistic
wave functions is discussed in Sec.7 in connection with the
so called Hegerfeldt theorem.

A short  discussion of  results  and  prospects
of refinement and development of the obtained structure is given in
the  concluding Sec.8.

 It should be noted at once that  an heuristic (or naive) level of
mathematical rigor is adopted and  a majority of  assertions of
are of general situation, that is  the necessary mathematical conditions
are supposed to be fulfilled. For example,  "asymptotic" means actually
"formal asymptotic" throughout the paper. I hope that it is plausible
because my first aim is to reveal possible changes in quantum
mechanics  related to or suggested by introduction of the Riemannian
geometry of space--time. A necessary mathematical refinement
can be made   if the primary results and  further development
of them	 prove to be interesting.

 Notation is standard  for general relativity
and, as a rule,  in the simple index form,  though, when it cannot cause a
confusion, indexless notation, like, e.g.,
$ \tau \nabla \equiv  \tau^\alpha \nabla_\alpha $,
will also  be used for brevity. {\it The dot  between differential
operators denotes  operator product} of them, i.e. $\hat A \cdot \hat B$
means  that $\hat A \cdot \hat B\, \psi(x) \equiv \hat A (\hat B\psi (x))$.\\

\noindent
{\large\bf 2. Quantization of Classical Mechanics in
Riemannian Space--Time}\\

   Quantization, according to  Dirac [12] ,  is  a linear map
$ Q: f \rightarrow \hat f $  of the Poisson algebra of functions
$ f \in C^\infty (M) $  on a symplectic manifold $ (M_{2n}, \omega) $,
$\omega$  being a symplectic form, to a set of operators acting
in a pre-Hilbert space  $\cal H$ ({\it the representation space}),
provided the following conditions are fulfilled:

1) $ 1\, \rightarrow \,\hat 1 $ ;

2) $ \{f , g\}_{\mbox{\footnotesize Poisson}}\quad \rightarrow
i\h^{-1}  [\hat f, \hat g]\
\stc\ i\h^{-1}  (\hat f \hat g - \hat g \hat f) $;

3) $ \ {\hat{\ov f}} = (\hat f)^\dagger $, where {\it
the dagger denotes the  Hermitean conjugation  with respect the scalar
product of $\cal H$};

4) a complete set $ \hat f_1, ..., \hat f_n $ of Hermitean operators exists,
such that,  if
 $\left[\hat f,\ \hat f_i \right] = 0  $ for any  $ i $, then
$\hat f = const \cdot \hat 1  $ .

   The map $Q$ cannot be found for an arbitrary  $M_{2n}$  but for
the dynamics of a point particle in \ri  a solution of  the problem
in the framework  of the  geometric quantization is presented in the
monograph by \'Sniatycki [11],  I have no  possibility
(and a capacity, too) to enter into details of the geometrical quantization
in the present paper. Instead, as a  primitive user,
I describe very briefly the initial  $M_{2n}$ and
resulting map $Q$  related to the geodesic dynamics in \ri following
to [11].

For a point--like particle moving along geodesic lines in \ri, the manifold
$M_{2n}$   is $ T^* \rif $, a cotangent bundle over \ri with a projection
$\ \pi: \,  T^* \rif \rightarrow \rif $.   Any  appropriate set
$\{ q^{(\alpha)} (x)\},\ x \in \rif $  of
four functions which satisfy the condition
$ \det\|\partial_\alpha  q^{(\beta)}(x)(\|  \neq 0$  defines
on $ T^* \rif $ a set of functions
$q^{(\alpha)} =  q^{(\alpha)} (x) \circ \pi$   which are constant
on fibers of  $ T^* \rif$ and will be referred following [11], as
{\it position type functions}. It is important to keep in mind  that
in the present  Sec.2 $q^{(\alpha)}$ and $q^{(\alpha)} (x)$ are different
functions: their domains are $ T^* \rif$, and \ri respectively.  Then,
a given chart  $\{U;\ x^0,\, x^1,\, x^2,\, x^3 \}$ in \ri  defines  on
a canonical chart
$$
\left\{\pi^{-1} (U),\ q^{(0)},...,\, q^{(3)}, p_{(0)},..., p_{(3)}\right\}
$$
on $ T^* \rif $ where  functions $p_{(\alpha)}$ are determined so
 that  $\omega = dp_{(\alpha)} \wedge dq^{(\alpha)} $ on $\pi^{-1} (U)$.

 An important point  for us in this construction is that,  on the background
of  the initial arbitrary curvilinear coordinates $\{x^\alpha\}$ which
provide $ U_A \subset \rif $ with an abstract arithmetization,    we
have introduced a set $ q^{(\alpha)} (x)$ of  four scalar functions
which is related to the phase space of the particle, may be quantized and
will be considered further as classical observable of space--time position
since the values of the functions also define a point on \ri .

  In the introduced notation  the general--relativistic dynamics of a
point--like particle of the rest mass  $ m $ on  $U$  is determined by the
constraint
\begin{equation}
 m^2 c^2
= p_{(\alpha)} p_{(\beta)} \left(g^{(\alpha)(\beta)} (x) \circ \pi\right),
\label{mc}  \\
\end{equation}
where
$$
 g^{(\alpha)(\beta)} (x) \bigr|_U
= \frac{\ptl q^{(\alpha)} (x)}{\ptl x^\gamma}
\frac{\ptl q^{(\beta)}(x)}{\ptl x^\delta}\ g^{\gamma\delta} (x).
$$
Thus, on the classical level,  the primary observables of the basic
physical interest and a constraint  on them  are
introduced. The resulting map $ Q $ of quantization  for these observables
can be exposed  according to [11], Sections 1.8, 10.1,   as follows.

The  representation space $\cal H$  is $L_2 (\rif,\,C) $,  a
 space of the complex valued square--integrable
 over \ri  and sufficiently smooth functions $\varphi (x)$. The
$L_2 (\rif,\,C) $ norm generates  an inner product in $\cal H $
determined as
\begin{equation}
 <\varphi_1, \varphi_2>
= \int_{\rif} \overline \varphi_1\, \varphi_2 \, dv_4	,
\qquad \varphi_1, \varphi_2 \in \cal H . \label{inn}
\end{equation}
$dv$ being the invariant volume element of \ri, i.e.
$$
dv\bigr|_U = (-g)^{1/2} (x) dx^0 dx^1 dx^2 dx^3,
\quad g(x) \stc \det \ g_{\alpha\beta} (x) \bigr|_U
$$

The operators $ \hat q^{(\alpha)}$ associated to the position  type
in  $ T^* \rif $ variables  $q^{(\alpha)}$
which may play the role of a complete set  of functions in the
condition  4) of quantization act on $\cal H$    as
\begin{equation}
\hat q^{(\alpha)}\bigr|_{\pi^{-1} (U)} \,  \varphi (x) =
q^{(\alpha)}(x)\, \varphi(x), \quad x\in U.    \label{qq}
\end{equation}
Thus  $ \hat q^{(\alpha)}$  form a complete commutative set
of operators and the condition  4) of quantization is satisfied.

Instead of  operators  $\hat p_{(\alpha)}$, it is convenient to  introduce
first  an operator  of projection of the momentum on  a given smooth vector
field $K^\alpha (x),\  x\in \rif\  $
\begin{equation}
\hat p_{K} (x)\bigr|_{\pi^{-1} (U)} \,
= i \hbar\ \left(K^\alpha (x) \nabla_\alpha
 + \frac{1}{2} \nabla_\alpha K^\alpha (x) \right).    \label{pks}
\end{equation}
where $\nabla_\alpha $ is {\it the covariant derivative} in \ri.
The operators
canonically conjugate to $\hat q^{(\alpha)}\bigr|_{\pi^{-1} (U)}$ are
given by the
 fields $ K^\gamma_{(\beta)}$ which are defined so that
$ K_{(\beta)}^{\alpha}  \ptl_\alpha  q^{(\gamma)} (x)
= {\delta^{(\gamma)}}_{(\beta)}$.

  The operators  $\ \hat q^{(\alpha)}, \quad \hat p_K $ are obviously
Hermitean with  respect to the inner product $ <.\ ,\ .> $.
Also, one   can  easily see that
\begin{equation}
\left[\hat p_K , \hat p_L\right] =
i \hbar\ \hat p_{[K,L]_{\mbox {\footnotesize Lie}}}, \label{kl}
\end{equation}
where $ [K,L]_{\mbox{\footnotesize  Lie}} $ is the Lie derivative
of the vector field $L$ along  $K$. Hence, there is  a commutative set
of four operators $ \hat p_{K_{(\alpha)}} $, since the vector fields
$ K_{(\alpha)}^\beta $ commute.

The constraint Eq (\ref{mc}) is mapped by $Q $  to  the condition
 specifying  in  $\cal H $   a subspace of  functions satisfying
the   equation
 \begin{eqnarray}
\Box\varphi + \zeta\, R(x)\, \varphi
+ \left(\frac{mc}{\hbar}\right)^2  \varphi &=& 0,
 \quad x\in \rif \label{r} \\
\Box \stc  g^{\alpha\beta}\nabla_\alpha \nabla_\beta, \qquad & &\nonumber
\end{eqnarray}
 with $\zeta = 1/6 $ and  not with $\zeta = 0$
as one might expect from  the viewpoint of minimality of the coupling
to gravitation.  This is just consistent with the result of [13, 14]
where it had been shown that $\zeta = 1/6 $  is
necessary for correct particle interpretation of the quantum theory of
the scalar field $\varphi (x) $ in \ri. (Despite of that in
[13, 14] only  QFT in de Sitter space--time was considered, nevertheless
the conclusion on necessity of $\zeta = 1/6$ is quite general,  as it was
indicated in [14].)

However, the presented  construction meets a serious difficulty in
physical interpretation.
It manifests in that  those solutions of Eq.(\ref{r}) which
can be considered  in QFT  in particular space--times \ri as
one--particle   wave functions
have a diverging $L_2 (\rif,\,C) $ norm because it  demands
 on $\varphi (x) $ to decrease in time--like directions.
The simplest  examples are  any superposition of the negative--frequency
solutions of in \eu and of the  analogous  solutions  in the De Sitter
space--time  obtained in [13, 14].
Such   property is not compatible with the physical idea of a particle as
a stable object in \eu.

It is clear that the roots of the divergence are in the choice of
 $ T^* \rif $ as initial  $M_{2n} $  and in the  symmetrical  treatment of
space and time coordinates. However, a moment of time,  contrary to the
space position, is not a property  of a particle.
 Therefore the considered scheme
of quantization  does not lead  to  any analog of the standard
quantum mechanics where the  time represents an evolution parameter.
Quantization of the geodesical dynamics,
 after some sort of the 1+3--foliation of \ri by   space--like
hypersufaces serving as configuration spaces enumerated by a time--like
parameter, would correspond better to this purpose, but apparently
it has not been  done.
However, I am not ready here to develop this version of quantization.
To my mind,  its assumed  result is contained in and covered  by
an alternative
field--theoretical approach to the construction of quantum mechanics
which starts with basic setting of quantum field theory in \ri.
Nevertheless, the exposed results of  geometric quantization
of the particle mechanics  suggests  how to treat curvilinear  coordinates
as observables covariantly. It is clear also   that,
in the 1+3--foliation formalism   and immediate quantization
of geodesical dynamics,  the canonical  operators
of a spatial coordinate  and  (\ref{kl})   of pojection of momentum
will have the same form of Eqs. (\ref{qq}) and (\ref{kl})
with the modification of $q$ and $K$ to the analogous
objects on the spatial sections sections of \ri, at least, in
the case when the sections are formed by the normal geodesic translation
of a given Cauchy hypersurface $\Sg$, see Sec.4.
\\

\noindent
{\large\bf 3. Quantum Field Theory  in  Riemannian Space--Time }\\

\noindent
{\bf 3.1 The Fock Representation  Spaces}\\

Now let us pass to the idea  that a structureless neutral particle is, in a
sense, a quantum of  the canonically quantized  {\it real }  scalar field
$\hat\varphi (x), \ x \in \rif$ satisfying Eq.(\ref{r}).
The general structure of   a Fock representation space
can be described as follows, see, e.g., [15, 16]

Consider in $\Phi_c\ = \ \Phi \otimes {\bf C}$, the
complexification of the vector space  $\Phi $ of real solutions  to
Eq.(\ref{r}), and  a subspace $\Phi_c^\prime \subset \Phi_c $ such that
\begin{equation}
    \Phi_c^\prime\ =\   \Phi^- \oplus  \Phi^+  \label{c}
\end{equation}
where $  \Phi^\pm  $ are supposed to be mutually complex conjugate
vectors spaces.
They selected so that  the  sesquilinear (i.e. linear for the second
argument and antilinear  for the first one) functional
\begin{equation}
\{ \varphi_1, \ \varphi_2 \}_\Sigma \stc i
\int_\Sigma d\sigma^\alpha (x)
\left(\ov {\varphi_1}(x)\ \ptl_\alpha \varphi_2 (x)\
- \  \ptl_\alpha \ov {\varphi_1}(x)\ {\varphi_2}(x)\right),
\label{spr}
\end{equation}
where
$\Sg = \{ x \in \rif;\ \Sg(x) =
const, \ \ptl_\alpha \Sg \ptl^\alpha\Sg > 0\}$
 is a Cauchy hypersurface in  \ri and $d\sigma^\alpha (x)$ is its
{\it normal volume element}, is positive (negative) semidefinite
on $  \Phi^-  \, \left( \Phi^+ \right). $
The value of the form does not depend on the choice  of the Cauchy
hypersurface $\Sg$ so far as $\varphi_1$ and $\varphi_2 $  both are
solutions of the field equation (\ref{r}). Therefore the form can be
considered as a scalar product in  $  \Phi^- $ providing the latter
with a pre-Hilbert structure.

Suppose further that   there is a  basis
$\{\varphi (x;\, A)\}$  in $\Phi^- $
enumerated by  a multi--index $A$ having values on  a set $\{A\}$
with a measure $\mu (A)$ and orthonormalized with respect to the inner
product Eq.(\ref{spr}), i.e.
\begin{equation}
\int_{\{A\}} d\mu(A)\ f(A)\ \{\varphi (.\,;\, A),\ \varphi (.\,;\, B)\}_\Sg  = f(B)
  \label{ort}
\end{equation}
for any function $f(A)$ on $\{ A\}$.
 (The assumption on existence of a basis  can be considered as
an auxiliary one to come to basic functionals
defined below by Eqs.(\ref{N1}), (\ref{pk2}) and (\ref{Q1})).
Then,  the quantum field operator is represented as
\begin{equation}
 \hat \varphi (x)
= \int_{\{A\}} d\mu(A) \left(c^+(A)\, \ov\varphi (x;\,A)\, +\,
c^-(A)\, \varphi (x;\,A)\right)
\equiv \hat\varphi^+ (x) + \hat\varphi^- (x),   \label{a+}
\end{equation}
with  the operators
$c^+(A)$ and $ c^-(A) $ of creation and annihilation of the field modes
$\varphi^- (x;\, A) \in \Phi^- $
(of {\it quasiparticles}), which  satisfy the canonical commutation relations
$$
[c^+(A),\,c^+(A')] = [c^-(A),\,c^-(A')] = 0,\
\int_{\{A\}} d\mu(A)\, f(A)\, [c^-(A),\, c^+(A')] = f(A')
$$
for any smooth function $f(A)$. They act  in the Fock space $\cal F$
with the cyclic vector $ |0> $ ({\it the quasivacuum}) defined by equations
\begin{equation}
 c^-(A)\, |0> = 0.  \label{vac1}
\end{equation}
Since the decomposition, Eq.(\ref{a+}), in general,
can be done by an infinite set of ways  and still there is no
reason to single out one of them, a question arises of physical
interpretation, at least, of a particular choice of the decomposition.
It seems natural to look for a decomposition in which the modes
might be interpreted as relativistic wave functions  of a particle.
However, then  a question arises on the meaning of the notion "a particle".
I shall return to these questions in Sec.4 and now, for time being,
continue with the introduced  arbitrary Fock spaces.\\

\noindent
{\bf 3.2.  Basic Operators of  Observables in  Fock Spaces}\\

Now  operators of observables acting in $\cal F$ should be introduced.
Having in mind  reconstruction of quantum mechanics  in the configurational
space,  it is natural to consider as basic operators of observables in QFT
the following ones.

{\it The operator of the number of modes (or of quasiparticles)} is defined
by straightforward generalization to \ri of  the  operator of the number
of spinless neutral particles  of the standard  QFT in \eu,  see, e.g., [1],
Chapter 7, Sec.3:
\begin{equation}
\hat{\cal N}(\hat \varphi;\,\Sg)
\stc  \{\hat\varphi^+ ,\, \hat\varphi^- \}_\Sg.   \label{N}
\end{equation}

{\it The operator of projection of the momentum of the field
$\hat \varphi (x)$
on a given vector field} $K^\alpha (x)$ is also a standard expression:
\begin{equation}
\hat {\cal P}_K (\hat\varphi ; \, \Sg ) \ = \  :\int_\Sg
d\sigma^\alpha \ K^\beta T_{\alpha\beta} (\hat\varphi):, \label{pk}
\end{equation}
where  the colons mean the normal product in expression between them and
 $T_{\alpha\beta}$ is the metrical energy--momentum tensor for
$\hat\varphi$, see  [13]:
\begin{eqnarray}
2\h^{-1}\, T_{\alpha\beta}(\hat\varphi)
&=&  \ptl_\alpha \hat\varphi\, \ptl_ \beta\hat\varphi +
\ptl_\alpha \hat\varphi\, \ptl_ \beta\hat\varphi -
  g_{\alpha\beta}\,\left(\ptl^\gamma \hat\varphi\, \ptl_\gamma\hat\varphi
+ \left(\frac{mc}{\h}\right)^2 \hat\varphi^2
+ \zeta R  \hat\varphi^2\right)\nonumber\\
& &\qquad - 2\zeta \left(R_{\alpha\beta} + \nabla_\alpha\nabla_\beta
- g_{\alpha\beta}\Box\right)\hat \varphi^2. \label{tem}
\end{eqnarray}
(The factor $\h$ is introduced in Eq.(\ref{tem}) from
considerations of the dimensions assuming the dimension of $\varphi$ is
that of the inverse length). It
is well known that $ \hat {\cal P}_K (\hat\varphi ; \, \Sg ) $ does not
depend on the choice of $\Sg$ if $K^\alpha $ satisfies the Killing equation
\begin{equation}
   \nabla_\alpha K_\beta + \nabla_\beta K_\alpha = 0  \label{kil}
\end{equation}
and thus defines an isometry of \ri. Then the condition of invariance of
the quasivacuum with respect to this symmetry is
\begin{equation}
   \hat {\cal P}_K (\hat\varphi ; \, \Sg )\, |0> = 0   \label{vac}
\end{equation}
and it distinguishes a particular class of decompositions (\ref{c})
reduced in \eu to the  unique decomposition for the linear envelops of
negative--  and positive--frequency  exponentials.

A less obvious point is  introduction  of   QFT--operators which
in appropriate cases will give   observables describing
the space localization of a quantum particle. Contrary to the
considered case of the momentum, the Lagrangian formalism of the
field theory does not provide with   classical prototypes of these
observables, what is natural.
However, in  QFT such an observable  can make sense and three corresponding
operators  can  be almost uniquely and covariantly introduced
if one recalls  the position type functions $ q^{(\alpha)} (x)$
introduced in Sec.2 and adopts the general structure of the operators
$\hat {\cal N}$ and $\hat{\cal P_K}$ introduced above. Then, if
one accepts the  point of view that, for the observables in
the NRQM, there exist some prototypes  in the
relativistic QFT, then the following line of reasoning seems to be
satisfactory to define these prototypes.

For a given Cauchy hypersurface $\Sg $, three  functions
$  q_\Sg^{(i)}(x), \quad i,j,... = 1,\, 2,\, 3,$ satisfying the conditions
\begin{equation}
\left. \ptl^\alpha \Sg \  \ptl_\alpha q_\Sg^{(i)}\right|_\Sg =  0 , \qquad
\mbox{rank}\left\|\ptl_\alpha   q_\Sg^{(i)}\right\| = 3.    \label{q}
\end{equation}
define a point on $\Sg$. Their restrictions on $\Sg$ can serve as internal
coordinates on it. They may be called spatial position type functions with
respect to $\Sg$.
It is natural to impose on the corresponding three QFT--operators
the following  conditions:
\begin{enumerate}
\item
They should be  real local quadratic functionals, like
$\hat {\cal P}_K (\hat\varphi ; \, \Sg )$, in the operators
$\hat\varphi^\pm (x)$
and linear functionals in  $q_\Sg^{(i)} (x) $ expressed as invariant
integrals over  $\Sg$.
\item
They should not contain derivatives of $  q_\Sg^{(i)} (x) $.
\item
They should lead to the operator of multiplication by the corresponding
argument of the wave function in the limit of the standard NRQM
(i.e. $c^{-1} = 0 $)  in the inertial frame of reference.
\end{enumerate}

These conditions lead apparently to the unique set of three operators on
$\cal F$ which will be called further {\it (spatial) position type
operators} (with respect to $\Sg$):
\begin{equation}
\hat{\cal Q}^{(i)}\{\hat\varphi;\, \Sg\} \stc i
 \int_\Sg d\sigma^\alpha\ q_\Sg^{(i)}(x)\
\left(\hat\varphi^+(x)\ \ptl_\alpha\hat\varphi^- (x)\
- \  \ptl_\alpha\hat\varphi^+ (x)\ \hat\varphi^- (x)\right) \label{Q2}
\end{equation}

If $\rif \sim E_{1,3},\ \Sg \sim E_3\ $ and
$q_{E_3}^{(i)}(x) \equiv x^i,\ $
$ x^i$ being Cartesian coordinates on $E_3$, this  operator coincides
with one of two versions of the position operators that  had been
considered by Polubarinov [17].
Actually, for reasons of causality, which are not correct from the point
of view adopted in the present paper, Polubarinov had preferred another
definition of the  Cartesian version of position operator. However,
along with some other unsatisfactory properties, the latter of his
definitions does not satisfy the third of the conditions formulated above.
Therefore I proceed with the definition Eq.(\ref{Q2}) which, in a certain
sense, leads to a generalization for \ri of the known Newton--Wigner
operator, see Sec.6. \\

\noindent
{\bf 3.3. Restriction to the One--Quasiparticle Subspace of a Fock Space }\\

 Let us consider  a one--quasiparticle state vector in ${\cal F}$
\begin{equation}
  |\varphi> \stc \{\varphi,\,  \varphi \}_\Sg^{-1/2}
\int_{\{A\}} d\mu (A)\ \tilde \varphi (A)\,  c^+(A)\, |0>,    \label{phi}
\end{equation}
determined by  a complexified  field configuration
\begin{equation}
\Phi^- \ \ni\ \varphi (x)
= \int_{\{A\}} d\mu (A)\ \tilde \varphi (A) \, \varphi (x;\, A)
\end{equation}
It is normalized, i.e.
\begin{equation}
<\varphi | \varphi> = 1
\end{equation}
because according to Eq.(\ref{ort})
\begin{equation}
\{\varphi,\,  \varphi \}_\Sg = \int_{\{A\}} d\mu (A)\ |\tilde \varphi (A)|^2
\end{equation}

 Consider now the matrix elements of operators
$\hat {\cal N} (\hat\varphi;\,\Sg),\
\hat {\cal P}_K (\hat\varphi; \, \Sg ) $
and $\hat{\cal Q}^{(i)}\{\hat\varphi;\, \Sg\}$ between  two such states
$|\varphi_1>$  and $|\varphi_2>$. Simple calculations  with the use of
Eqs.(\ref{N1}), (\ref{tem}), (\ref{Q2}) and (\ref{phi}) give
\begin{equation}
  <\varphi_1|\,  \hat {\cal N} (\hat \varphi;\,\Sg)\, |\varphi_2>
= \frac{\{\varphi_1,\, \varphi_2 \}_\Sg}
{\{\varphi_1,\, \varphi_1 \}_\Sg^{1/2}
\{\varphi_2,\, \varphi_2\}_\Sg^{1/2}} , \label{N1}
\end{equation}
\begin{equation}
<\varphi_1| \hat {\cal P}_K (\hat\varphi; \Sg)|\varphi_2> =
\frac{P_K (\varphi_1,\, \varphi_2;\, \Sg)}
{\{\varphi_1,\, \varphi_1 \}_\Sg^{1/2}
\{\varphi_2,\, \varphi_2\}_\Sg^{1/2}} \label{pk1}
\end{equation}
where
\begin{eqnarray}
P_K (\varphi_1,\, \varphi_2;\, \Sg) = \h\int_\Sg \, d\sigma^\alpha \,
 \biggl(\ptl_\alpha \ov\varphi_1\, K^\beta\ptl_\beta \varphi_2
+ K^\beta \ptl_\beta\ov\varphi_1\, \ptl_\alpha \varphi_2
\qquad\qquad\qquad\qquad\qquad\qquad & & \nonumber \\
\qquad  -\,  K_\alpha
\left(\ptl_\beta \ov\varphi_1\, \ptl^\beta \varphi_2 -
\left(\left(\frac{mc}{\hbar}\right)^2
+ \zeta R\right)\ov\varphi_1\,\varphi_2\right) -
\zeta\, K^\beta \left(R_{\alpha\beta}
+ \nabla_\alpha\nabla_\beta
- g_{\alpha\beta}\Box\right)\, (\ov\varphi_1 \varphi_2) \biggr), \label{pk2}
\end{eqnarray}
and
\begin{equation}
<\varphi_1|\,\hat{\cal Q}^{(i)}\{\hat\varphi;\, \Sg\}\,|\varphi_2> =
\frac{\{\varphi_1,\, q_\Sg^{(i)} \varphi_2 \}_\Sg}
  {\{\varphi_1,\, \varphi_1 \}_\Sg^{1/2}\{\varphi_2,\, \varphi_2\}_\Sg^{1/2}}
\label{Q1}
\end{equation}
The right--hand side of Eq.(\ref{pk2}) can be simplified by subtraction
of the divergence $\nabla^\alpha S_{\alpha\beta}$ of an
antisymmetric tensor $S_{\alpha\beta}$  from the integrand, which does
not contribute to the integral according to the Gauss theorem.  Taking
$$
S_{\alpha\beta} \stc \zeta \left(K_\alpha \ptl_\beta - K_\beta \ptl_\alpha
+ \frac12  (\nabla_\alpha K_\beta - \nabla_\beta K_\alpha)\right)\,
(\ov\varphi_1 \varphi_2),
$$
one obtains for the last three terms  in Eq.(\ref{pk2})
\begin{equation}
\int_\Sg \, d\sigma^\alpha \,\zeta K^\beta \left(R_{\alpha\beta}
+ \nabla_\alpha\nabla_\beta
- g_{\alpha\beta}\Box\right)\, (\ov\varphi_1 \varphi_2)
= \int_\Sg \, d\sigma^\alpha\, \zeta (\tilde K_{\alpha\beta} \ptl^\beta -
\nabla^\beta \tilde K_{\alpha\beta})\,(\ov\varphi_1 \varphi_2)
\end{equation}
where
\begin{equation}
\tilde K_{\alpha\beta} \stc \nabla_\alpha K_\alpha +
\nabla_\beta K_\alpha - \nabla K \, g_{\alpha\beta}
\end{equation}
and $\nabla K \stc \nabla_\gamma K^\gamma $ . The tensor
$\tilde K_{\alpha\beta}$  evidently vanishes when $ K_\alpha$ is
a Killing vector and  ${\cal P}_K  (\varphi; \, \Sg)$ is
a conserved quantity  (that is it does not depend on  the choice of $\Sg$ ).

Eqs.(\ref{N1})--(\ref{Q1})  for the matrix elements  of the basic observables
in the one--quasiparticle subspace of a given  $\cal F$
reveal  a projective structure in the space $\Phi^-\ $:  $\ \varphi(x) $ and
$const \cdot \varphi(x)$ are equivalent one--quasiparticle wave functions
for calculation of matrix elements
$<\varphi_1|\,\hat{\cal N}\{\hat\varphi;\, \Sg\}\,|\varphi_2>$,
 $\ <\varphi_1|\,\hat{\cal P}_K\{\hat\varphi;\, \Sg\}\,|\varphi_2>$ and
$<\varphi_1|\,\hat{\cal Q}^{(i)}\{\hat\varphi;\, \Sg\}\,|\varphi_2>$
though they are different as  complex superpositions of
 real classical scalar fields.

The sesquilinear functionals  $\{\varphi_1,\, \varphi_2 \}_\Sg$,
$P_K (\varphi_1,\, \varphi_2;\, \Sg)$ and
$\{\varphi_1,\, q_\Sg^{(i)} \varphi_2 \}_\Sg$   of
$\varphi_1 (x),\  \varphi_2 (x)\in \Phi^- $
are obviously Hermitean  in the sense that,  given  a functional
${\cal Z}(\varphi_1,\,\varphi_2;\, \Sg)$, the following equality takes place:
\begin{equation}
{\cal Z}(\varphi_1,\,\varphi_2;\, \Sg)
= \ov{{\cal Z}(\varphi_2,\,\varphi_1;\, \Sg)}.  \label{z}
\end{equation}
They determine  quantum mechanics of a quasiparticle specified by
decomposition (\ref{c}) provided that the processes  of creation and
annihilation of the quasiparticles by the gravitational field
can be neglected.

In principle, one could   proceed further with
$\{\varphi_1,\, \varphi_2 \}_\Sg$,
$P_K (\varphi_1,\, \varphi_2;\, \Sg)$ and
$\{\varphi_1,\, q_\Sg^{(i)} \varphi_2 \}_\Sg$
as quantum--mechanical amplitudes  of transition under measurement
of the corresponding observable. S.Weinberg [18] formulated  a version
of NRQM with a nonlinear \Sche in terms of
an open algebra of Hermitean  functionals of observables. An extension
of  Weinberg's formalism to our case seems possible in principle and
even useful for consideration  of quantum mechanics in
essentially  different  frames of reference but this question
needs a special study. Here I shall  develop the traditional
operator formalism. \\

\noindent
{\large\bf 4. Asymptotic Quasinonrelativistic  One--Particle
Wave Functions}\\

Now  the main  problem is to distinguish  that  space $\Phi^-$
 which could be interpreted on sufficient physical basis
as the space of wave functions of particles instead of the ambiguous
notion of quasiparticles. In  \eu and globally static space-times (see
definition in Sec.6) there  exists  a  unique decomposition Eq.(\ref{c})
such that  an irreducible representation of the
space-time symmetry is realized on  $\Phi^- $.
Such distinguished Fock spaces
are singled out also in the de Sitter and Friedman--Robertson--Walker
nonstationary cosmological models but
only by  combination of the symmetry arguments with  additional
physical arguments such as a correct   quasiclassical behavior  of
$ \varphi(x) \in \Phi^- \ $  [13, 19],
 minimality of the rate of cosmological  particle creation [20],
diagonalization of the field hamiltonian [21].

In the general \ri  one has no symmetry arguments and
can  appeal only to an intuitive idea of a quantum particle
as a localized  object, which  is firmly formulated only in the
Schr\"odinger representation of the standard nonrelativistic quantum
mechanics where  $c^{-1} = 0$. A choice of the Fock space realizing
this idea can be done only by an immediate
construction of a space $\Phi^-$, which, in turn, can be done in
the general  \ri only by approximate methods. Having in mind the
standard NRQM as a guideline, I shall consider   the space
$ \Phi^- (\Sigma;\, N)$ of formal asymptotic solutions of Eq.(\ref{r})
of an order N in $ c^{-2} $ of  the following  WKB--type form:
\begin{equation}
\varphi (x) \ = \ \sqrt{ \h/2mc }\
exp\left(-i \frac{mc}{\h}\ S_\Sg(x)\right)\ \phi (x).  \label{az}
\end{equation}
The function $S_\Sg (x)$ is assumed to be a solution of
the Hamilton--Jacobi equation
\begin{equation}
\ptl_\alpha S_\Sg \,\ptl^\alpha S_\Sg \ =\ 1,  \label{hj}
\end{equation}
 with the initial value  $\ S_\Sg (x)	\mid_\Sg = S_0 \equiv const.$
on an initially fixed  Cauchy hypersurface  $\Sigma $.
Thus, any  hypersurface
$\ S_\Sg(x) =  const\ $ forms a  level surface of
a geodesic flow normal to $\Sg $. These hypersurfaces
which will be denoted further as $S_\Sg  $ or simply as $S$ may be called
{\it a normal geodesic translation $S_\Sg $  of a  given}  $ \Sigma $.
Thus an 1+3--foliation  of \ri is introduced and the  value of
$ S_\Sg  $ at given $x\ \in \rif $  can be considered as
 an evolution parameter. Of course, this  is only a covariantization
introduction of the semigeodesic, or Gaussian, coordinates in \ri.

   Until now , the matter resembles the quasiclassical approximation for
a second order  differential  equation with  small parameters
$ \h^2, \ c^{-2},\ \mbox{or}\  m^{-2}$  at second derivatives, see, e.g.
[22].  One goes to {\it the quasinonrelativistic
approximation} instead of the quasiclassical one   when one introduces
{\it a normal geodesic frame of reference},  i.e. a time-like vector field
 \begin{equation}
\tau^\alpha\  \stc\  c\ \ptl^\alpha S_\Sg  ,\qquad
\tau^\alpha \tau_\alpha\ = \ c^2  .
\end{equation}
This is equivalent to the introduction of a variable $t\ = \ c^{-1} x^0$ of
the dimensionality of the  macroscopic time, after which the left--hand
side of Eq.(\ref{r}) ceases to be a $\lambda$--differential operator
in terms of [22] with  $ \lambda = c^{-1}$.

Then, if Eq.(\ref{az}) is an asymptotic solution of Eq.(\ref{r}), i.e.
 \begin{equation}
\Box\varphi + \zeta R(x)\, \varphi  +
\left(\frac{mc}{\hbar}\right)^2  \varphi = \Oc,
 \label{ar}
\end{equation}
one comes through  obvious iterations to the following evolution
equation for $\phi (x)$
\begin{equation}
i\h\ T\ \phi (x) \ = \ H_{N}\ \phi (x), \quad
\ H_N\ \stc  H_0\ +\ \sum_{n=1}^N \frac{h_n}{2mc^2} + \Oc, \label{t1}
\end{equation}
where
\begin{equation}
T \ \stc \ \tau^\alpha\nabla_\alpha\
+\ \frac12 \nabla_\alpha \tau^\alpha, \label{t0}
\end{equation}
cf. Eq.(\ref{pks}), and
\begin{equation}
H_0 \ \stc \ -  \frac{\h^2}{2m} \left( \triangle_{S_\Sg}  - \zeta R +
\left(\frac{1}{2}(\ptl S_\Sg\ \ptl\Box S_\Sg) \
+\  \frac{1}{4} (\Box S_\Sg)^2 \right)\right). \label{h0}
\end{equation}
$\triangle_{S_\Sg} (x) $ is the Laplace--Beltrami operator on the
hypersurface $S_\Sg (x) = const$.   The differential operators $h_n  $
are determined by  recurrence relations:
\begin{equation}
  h_{n+1} \ =\ [-i\hbar\ T, \ h_n]\ -\ \sum_{k=0}^{n}\ h_k\cdot h_{n-k},
\quad n > 0; \quad\ h_0 \equiv H_0 .  \label{hn}
\end{equation}

It is easy to see that the differential operator $ H_N \equiv H_N (x) $
which will be further the main element of construction  contains
only {\it covariant derivatives $D_\alpha$ along the hypersurface $S$ }
to which the point $x$  belongs:
\begin{equation}
 D_\alpha \stc h_\alpha^\beta \nabla_\beta, \quad
h_{\alpha\beta} \ \stc\ c^{-2}\, \tau_\alpha  \tau_\beta\
- \ g_{\alpha\beta}, \label{d}
\end{equation}
i.e. $ h_{\alpha\beta}$  is {\it the tensor of projection on  $ S $}.
For example, $\triangle_S\ =\ -D^\alpha  D_\alpha$.

Now, turn to the  scalar product   $\{\varphi_1,\, \varphi_2 \}_\Sg,\ $
Eq.(\ref{spr}), on  $\Phi^- (\Sg; \, N) $, which, according to Eq.(\ref{N1})
is  a matrix element of $\hat {\cal N} (\hat\varphi;\,\Sg)$,
the  QFT--operator of the number of quasiparticles.  It is obvious  that
\begin{equation}
\{\varphi_1,\, \varphi_2 \}_{S_\Sg} = \{\varphi_1,\, \varphi_2 \}_\Sg + \Oc.
   \label{ssg}
\end{equation}
and thus  the quasiparticle is asymptotically stable in the neighborhood of
$\Sg$, in which the accepted approximation is valid.

The inner product $\{\varphi_1,\, \varphi_2 \}_\Sg $ is asymptotically
positive definite  on $\Phi^- (\Sg; \, N)$ in the sense  that,
for  $ g_{\alpha\beta} \in  C_{2N} (S)$,
a sufficiently small value of $c^{-2} > 0$ exists for which
$\{\varphi_1,\, \varphi_2 \}_\Sg $  is positive definite.  In a
physical sense this is, of course, a condition on the metric, the
function  $\varphi (x) $ and on their derivatives.
Thus,  it induces an asymptotic positive definite norm
$\{\varphi,\, \varphi \}_\Sg^{1/2}$  which, however, is not an
$L_2 (\Sg;\, C)$ norm which would be a natural generalization
of the $L_2 (E_3;\,C)$ norm
of the standard  NRQM in the \Schr. The latter norm is essential
in the quantum mechanics for a precise definition  of localization of
a particle in terms  of the projection--valued measure on  $E_3$ in
Cartesian coordinates, see [23],  Sec.13--1.

  More simply speaking, the integrand of $\{\varphi,\, \varphi \}_\Sg$
is not nonnegative and therefore can not be interpreted as a probability
density on $\Sg$.
( For \eu,  an example of a  superposition of positive--frequency
exponentials  for which the integrand oscillates between positive and
negative values  can be found in [24].) If it were positive everywhere on
$\Sg$ then one could  restrict  the integration
in expression  for $\{\varphi,\, \varphi \}_\Sg$ to any
domain  $\triangle\Sg \subset \Sg $  and consider this modified
quadratic functional as the probability of
detecting a particle  in $\triangle\Sg$, what would correspond
to a Born probabilistic interpretation of $\varphi$.
Then one might restrict to $\triangle\Sg$ the
integrals in expressions for
$P_K (\varphi,\, \varphi;\, \Sg)$  and
$\{\varphi,\ q^{(i)}_\Sg \varphi \}_\Sg$
and, substituting
$P_K (\varphi,\, \varphi;\, \triangle\Sg )$  and
$\{\varphi,\ q^{(i)}_\Sg \varphi \}_{\triangle\Sg}$ thus obtained into
Eqs.(\ref{pk1}) and (\ref{Q1}) for
$<\varphi| \hat {\cal P}_K (\hat\varphi; \Sg)|\varphi >$  and
$<\varphi|\,\hat{\cal Q}^{(i)}\{\hat\varphi;\, \Sg\}\,|\varphi >$,
come to average values of these observables on $\triangle\Sg$.

For a free motion in \eu,  a mapping of $\Phi^- $ to a space
with the  $L_2 (E_3;\, C)$   is given by
the Feshbach -- Villars transformation originally set in
the momentum representation [1]. The presence of an external field
forces to look for a similar transformation  in the configurational
representation.  Therefore,  I consider $\phi (x) $ (and, consequently,
$\varphi (x)$) as an asymptotic transformation of another function
$\psi (x)$:
\begin{equation}
\phi (x) \ =\ V_N (x,\, D)\ \psi(x), \qquad \psi (x)\mid_S \,
\in L_2(S;C)\ \mbox  {for any}\ S \label{v}
\end{equation}
and define the asymptotical differential operator $V_N (x, D)$
which acts along the
hypersurface $S$ containing the point $x \in\rif $ so that
the following  relation takes place:
\begin{equation} \left(\psi_1,\ \psi_2\right)_S \ \stc
\ \int_S d\sigma(x) \ \ov\psi_1\ \psi_2 \
=  \{\varphi_1,\ \varphi_2 \}_S +  \Oc, \label{psi}
\end{equation}
$d\sigma(x)$ being {the invariant volume element} of $S$. Hence  and from
Eq.(\ref{t1}) it follows that $V_N$  satisfies  up  to multiplication from
the right by an arbitrary unitary differential operator the equation
\begin{equation}
V_N\cdot V_N^\dagger \  = \
\left(1 \  + \ \frac{ H_N + H_N^\dagger}{2mc^2} \right)^{-1} \
+ \ \Oc.\label{vv}
\end{equation}
Here and further  the Hermitean conjugation denoted by the dagger
is defined  with respect to the scalar product
$\left(\psi_1,\ \psi_2\right)_S$,  that  is, for example,
\begin{equation}
\left(H \psi_1,\ \psi_2\right)_S\
= \ \left(\psi_1,\ H^\dagger\psi_2\right)_S .
\end{equation}

It is obvious that Eq.(\ref{v})
 defines $V_N$  up  to multiplication from the right by an
arbitrary asymptotically unitary differential operator.

It is easily seen   from Eq.(\ref{t1}) that  $ \psi (x) $ satisfies
the  following  Schr\"odinger equation:
\begin{equation}
i\h\ T\ \psi \ = \ \hat H_N \ \psi, \label{t2}
\end{equation}
where
\begin{equation}
\hat H_N \stc \ V_N^{-1}\cdot \left(H_N\cdot V_N\ +\ [i\h T,\ V_N]  \right).
 \label{ham}
\end{equation}

Further, any differential operator $\hat z(x, D)  $
which is a polynomial of the order $2N$ of the "spatial" derivatives
$D_\alpha$, Eq.(\ref{d}), the
commutator $[i\h T,\ \hat z]$, being restricted to  the
space   of solutions  of Eq.(\ref{t2}),
is again such a polynomial.  Hence, taking into account  the relation
\begin{equation}
[i\h T,\ \hat z]^\dagger \ =\ -\ [i\h T, \ \hat z^\dagger],
\end{equation}
which is not so obvious because there is no Hermitean conjugation
for operator $T$ (see [8]  for the  proof of the relation ),
one can see that the hamiltonian operator $\hat H_N $,
in contrast to $H_N $,  is asymptotically Hermitean, i.e.
\begin{equation}
\hat H_N\ =\hat H_N^\dagger \ +\ \Oc.   \label{ham1}
\end{equation}
  Therefore  the sesquilinear form
 $\left(\psi_1,\, \psi_2 \right)_S$ does not depend on the
value of S (though depends on the choice of the initial $\Sg$
which generates the 1+3--foliation ) and can be considered as
a scalar product in {\it the space $\Psi (\Sg;\, N) $ of solutions of the
\Sche} Eq.(\ref{t2}). Then  $|\psi(x_1)|^2\, d\sigma (x_1)$  can be
considered as  a  density  of probability  to observe the
asymptotically stable configuration
described by {\it the quasinonrelativistic wave function} $\psi (x)$,
 or, equally, by the corresponding $\varphi (x)$, at the point $x_1$
of the hypersurface $S(x) = S(x_1)$. \\

\noindent
 {\large\bf 5. Quasinonrelativistic Operators of Observables in \\
 the  Field--Theoretical Approach}\\

Having accepted  the point of view that  $ \Psi(\Sg;\, N) $ is
the projective space  of states of a quantum spinless  particle in \ri
with   the Born's  probabilistic  interpretation of $\psi(x) $ one
should  introduce  a way to evaluate  mechanical observables of
the particle in the state defined by $\psi(x) $. Now  I shall
do it on the same field--theoretical basis.  \\

According to Eqs.(\ref{N1}), (\ref{ssg}), (\ref{psi})
\begin{equation}
<\varphi_1|\,  \hat {\cal N} (\hat \varphi;\, S )\, |\varphi_2>
= \frac{\left(\psi_1,\, \psi_2\right)_S}
{\left(\psi_1,\, \psi_1\right)_S^{1/2}
\left(\psi_2,\, \psi_2\right)_S^{1/2}} + \Oc,  \label{n3}
\end{equation}
that is the operator of number of particles
$ \hat {\cal N} (\hat \varphi;\,\Sg) $
is represented in  $ \Psi(\Sg;\, N) $  by the unity operator as
it should be in the quantum mechanics of a single stable particle.\\

\noindent
{\bf 5.1. Quasinonrelativistic Operators of Momentum
and Energy of a Particle}\\

  Like $\hat {\cal N}$,  the one-particle matrix element
(\ref{pk}) of the QFT--operator of the  projection of momentum
can be represented as a matrix element of an differential  operator
  $\hat p_K (x;\, S;\, N) $  {\it acting along $S$ on $ \Psi(\Sg;\, N) $,
i.e. containing only "spatial" derivatives $D_\alpha$}:
\begin{equation}
<\varphi_1| \hat {\cal P}_K (\hat\varphi; S)|\varphi_2> =\
\frac{ \left(\psi_1,\, \hat p_K (.\, ;\, S; \, N)\, \psi_2\right)_S}
{\left(\psi_1,\, \psi_1\right)_S^{1/2}
\left(\psi_2,\, \psi_2\right)_S^{1/2}} + \Oc,      \label{pk3}
\end{equation}
Obviously,
\begin{equation}
p_K (x;\, S;\, N) = p_K^\dagger (x;\, S;\, N) + \Oc
\end{equation}
owing to the property (\ref{z}) of
$<\varphi_1| \hat {\cal P}_K (\hat\varphi; S)|\varphi_2>$.
It is natural to consider  the operator
$\hat  p_K (x\, ;\,S_\Sg;\, \, N) $ as
{\it the quasinonrelativistic operator of the projection of momentum} on
a given vector field $K^\alpha$ in \ri, that is as an  analog of the
momentum operator of the standard NRQM, but now $c^{-1} \neq 0$.

A straightforward calculation with the use of properties
of $\tau^\alpha $, of the relation
$$
D_\alpha^\dagger = -D_\alpha - c^{-2} \tau_\alpha\,\nabla_\beta\tau^\beta ,
$$
 and of Eqs. (\ref{t1}), (\ref{v}), (\ref{vv}) gives
\begin{eqnarray}
& &{\hat p}_K (x;\,S;\,N)
 =  \frac12\  V_N^{\dagger}\cdot \biggl\{ m\, \tau K
+  \left( 1 + \frac{\tilde H^{\dagger}_N}{mc^2}\right)\cdot
m\,\tau K^
\cdot \left(1 + \frac{\tilde H_N}{mc^2}\right)\qquad\qquad \nonumber\\
& &\qquad
- \left(1 + \frac{\tilde H^{\dagger}_N}{mc^2}\right)\cdot i\h K D
- (i\h K D)^{\dagger}\cdot  \left( 1 + \frac{\tilde H_N}{mc^2}\right)
- \frac1{mc^2} \left( \left(i\h D_\alpha\right)^{\dagger}\cdot
\tau K\cdot i\h D^\alpha \right)  \nonumber\\
& &\qquad +  \frac{\zeta\h}{2mc^2} \left(\frac{i}{c^2}
\left( (\tau \tilde K \tau) \cdot \tilde H_N
- \tilde H_N^\dagger\cdot (\tau \tilde K\tau)\right)
- \h W(K)\right) \biggr\}\cdot V_N +  \Oc,
\end{eqnarray}
where
$$\tilde H_N \stc  H_N - \frac12 i\h \nabla\tau\, $$
 an indexless notation like
\begin{equation}
\nabla\tau \stc \nabla_\beta\tau^\beta,\
(\tau \tilde K \tau) \stc \tau^\alpha \tilde K_{\alpha\beta} \tau^\beta
\end{equation}
is used for simplicity, and
\begin{equation}
W(K) \stc D^\alpha(\tau^\beta \tilde K_{\alpha\beta})
- \tau^\beta \nabla^\alpha \tilde K_{\alpha\beta}.
\end{equation}

There are two explicitly distinctive  samplings of $K^\alpha$:
$ \tau_\alpha K^\alpha \equiv \tau K = 0$  and
$  K^\alpha = c^{-1} \tau^\alpha $.
In the first case one obtains the spatial  projection  of  momentum:
\begin{eqnarray}
\hat p_K (x;\,S; N)\mid_{\tau K=0}  = -\frac{1}{2}\ V^\dagger_N \cdot
\left(\left( 1 + \frac{\tilde H^\dagger _N}{mc^2}\right)\cdot i\h K D \,
+\, (i\h K D)^\dagger
\cdot \left(1 + \frac{\tilde H_N}{mc^2}\right)\right.& & \nonumber\\
- \frac{\h\zeta}{2mc^2}\bigl(i\,\nabla K\cdot H_N
- i\, H_N^\dagger \cdot\nabla K
- \h W(K)\bigr) \Biggr)\cdot V_N + \Oc.& & \label{psp}
\end{eqnarray}

For $N =1$  and
$ V_1^\dagger =  V_1 $ Eq.(\ref{psp}) takes the form
\begin{eqnarray}
{\hat p}_{K}(x;\,S;\,1)\mid_{\tau K = 0}
& = & i\h \left(K \nabla + \frac12 \nabla K \right) +
\frac{\h}{4mc^2}\,(D\cdot K)\nabla\tau   \nonumber\\
& + &\frac{1-2\zeta}{4 mc^2}\,[i\h\nabla K,\, H_0]
- \frac{\h^2\zeta}{2 mc^2} W(K)  +  O\left(c^{-4}\right) \label{psp1}
\end{eqnarray}

We see that for $N=0$, i.e. for exact nonrelativistic limit,
this operator coincides with
$\hat p_{K}\mid_U (x)  $, Eq.(\ref{pks}).  It is
remarkable, however, that if $K^\alpha$ and $ L^\alpha $ are
two Killing vectors fields along the level hypersurfaces
$S$,  then Eq.(\ref{kl}) takes place for any value of $N$ owing to
the relation $[\tau \nabla,\, K D]~= [\tau \nabla,\, L D] = 0$.
Thus, the operators
$\hat p_{K_a} (x;\,S; N)$  as well as $\hat p_{K_a}\mid_U $ realize  a
representations of the Lie
algebra  of the group of isometry defined by the Killing fields
along surfaces $ S $. (They are also  Killing vectors of
the interior geometry
of $\Sg$ induced by the metric of \ri.) In particular, this means
that in the Friedman--Robertson--Walker space--times  there is
a complete set of
commuting  asymptotic operators defined by their spatial symmetries,
namely, $SO(3)$ for the closed model, $SO(1,2)$ for the open model and
$E(3)$ for the spatially flat model.

In general, we come here to a very interesting topic of representation
of Lie algebras  by asymptotic  operators, but it needs a special study.

In the case of $ K^\alpha = c^{-1} \tau^\alpha$ simple transformations
give  the operator of energy
\begin{eqnarray}
&& c\ {\hat p}_{\tau/c}(x;\, S;\, N)  =  mc^2
+ V_N^\dagger
\cdot \biggl(H_0  + \frac{\tilde H_N^\dagger\cdot \tilde H_N }{2 mc^2}
\nonumber\\
& &\quad \quad - \frac{\h\zeta}{2mc^2}\,([i\h T - \tilde H_N,\ \nabla\tau]
 -  \h (\nabla\tau)^2
-\h R_{\alpha\beta}\tau^\alpha\tau^\beta )\biggr)\cdot V_N +  \Oc.
\label{E}
\end{eqnarray}

Again for $V_1^\dagger = V_1  $\  one has
\begin{eqnarray}
c\ {\hat p}_{\tau/c}(x;\,S;\,1)\, &=& \, mc^2\,
 +  H_0\ - \frac{ H_0^2 }{2 mc^2}
  +  \frac{i\h}{4 mc^2}\, [(\nabla \tau),\ (1-2\zeta)H_0
-  2\zeta i \h \nabla\tau] \nonumber\\
 &+& \frac{1+ 4\zeta}{8mc^2}\,\h^2\, (\nabla \tau)^2
+ O\left(c^{-4}\right).
\end{eqnarray}

 A  very important point is that
the energy  operator $c\ {\hat p}_{\tau/c}(x;\, S;\, N)  $  is
unitarily equivalent to the hamiltonian $\hat H_N $ in \Sche  (\ref{t2})
in the sense that the operator $V_N \ $, having been defined by
Eq.(\ref{vv}) up to multiplication by an   asymptotically unitary
operator from the left, can be chosen so that the following equality will
take place  on $\Psi(S;\, N)$:
\begin{equation}
c\ {\hat p}_{\tau/c}(x;\, S;\,N)\   = \ mc^2\ + \ \hat H_N \ +\ \Oc.
 \label{u}
\end{equation}
The proof of this fact which is an important indication of the
self--consistency of the approach  is given in Appendix. \\

\noindent
{\bf 5.2. Quasinonrelativistic Operator of Spatial Position
of a Particle}\\

 A normal 1+3--foliation of \ri by the one -parametric  set
of hypersurfaces $S_\Sg$ having been done,   the position
type functions $q^{(i)}_\Sg (x)$ of Subsec 3.2. can be introduced
so that the conditions
(\ref {q}) are satisfied on each  $S_\Sg$. Thus, they are constant on
each geodesic which is normal to  $\Sg$  and translate an interior
coordinate system of $\Sg$ to each $S_\Sg$.

Then, similarly to the cases of  operators $\hat{\cal N}_\Sg $  and
$\hat {\cal P}_K (\hat\varphi; \Sg)$  a  spatial
operator of position which is Hermitean in $\Psi_N$  is defined
by the equality of  matrix elements:
\begin{equation}
<\varphi_1|\,\hat{\cal Q}^{(i)}\{\hat\varphi;\, S\}\,|\varphi_2>
= \frac{(\psi_1,\,\hat q^{(i)} (.\,;\, S;\, N)\,\psi_2)_S}
{\left(\psi_1,\, \psi_1\right)_S^{1/2}
 \left(\psi_2,\, \psi_2\right)_S^{1/2}} + \Oc \label{qs}
\end{equation}
from which it follows that
\begin{equation}
\hat q^{(i)} (x,\,S; N)\, = \, V_N^\dagger\cdot \left(q_S^{(i)} (x)\
+\ \frac{H_N^\dagger\cdot q_S^{(i)} (x)\,
+\, q_S^{(i)} (x) H_N}{2 mc^2} \right)\cdot V_N\, +\, \Oc.
\end{equation}

For $ V_2^\dagger =  V_2 $\
\begin{eqnarray}
\hat q^{(i)} (x,\, S;\, 2) &=& q_S^{(i)} (x)\
- \ \frac{1}{(2 mc^2)^2} \left(\bigl[ i\h T -
\frac{1}{2} H_0,\ [H_0,\ q_S^{(i)} (x)]\bigr]\right) \
+ O\left(c^{-6}\right) \nonumber \\
&=& q_S^{(i)} (x)
+ \ \frac{1}{(2 mc^2)^2}
\left(\frac{\hbar^2}{m} \hat p^0_{[\tau, \ptl q^{(i)})]}
+ \frac{i\hbar}{2m} \bigl[H_0,\ \hat p^0_{ \ptl q^{(i)}}\bigr]\right) +
O\left(c^{-6}\right), \label{qp}
\end{eqnarray}
where $ \hat p^0_K \stc \hat p_K (x,\, S;\, 0)$ and
$[\tau, \ptl q^{(i)}] \equiv [\tau, \ptl q^{(i)}]_{\mbox{\footnotesize Lie}}$,
the latter expression in Eq.(\ref{qp})  is a consequence of Eq.(\ref{kl}) and
of the relation
$$
\bigl[H_0,\  q_S^{(i)}(x) \bigr] = \frac{i\hbar}{m} \hat p^0_{ \ptl q^{(i)}}.
$$
It is remarkable that  the first relativistic correction vanishes and the
operators of space position type functions commute up to
$O\left(c^{-4}\right)$
and may be taken up to this accuracy as a complete  set of operators
of the observables, but for $N > 1$ they are noncommutative. Thus,
the field-theoretically determined operators of the space position
$ \hat q^{(i)}(x;\, S;\, N) $ and of the space momentum
$ \hat p_{\ptl q^{(i)}} (x;\, S;\, N) $
cannot coincide with the  canonically conjugated primary operators
of quantized mechanics unless $N=0$, i.e. except
the exact nonrelativistic limit.\\

\noindent
{\large\bf 6. Quasinonrelativistic Operators of
Observables in Globally Static \\  and Minkowskian Space--Times}\\

Consider now a globally static \ri where
a normal frame of reference  $\tau^\alpha (x) $ exists that satisfies
the Killing equation
$\nabla_\alpha\tau_\beta + \nabla_\beta \tau_\alpha = 0$.\
It means that $\tau^\alpha $ is a covariantly constant vector field.
Then, if $S(x) $ is chosen so that
 $ \tau_\alpha = c\, \ptl_\alpha S $, one has  $[T,\ H_0] = 0 $
and, having taken  $ V_N^\dagger =  V_N $,
comes to  the following  formal  closed expressions for
 $ N \rightarrow \infty $ :
\begin{eqnarray}
H_{\infty} = \hat H_{\infty} &=&  mc^2 \left(\left(1 +
\frac{2H_0}{mc^2}\right)^{1/2} - 1 \right),
\qquad  H_0 = -\frac{\h^2}{2m}(\triangle_S - \zeta \,R) , \label{hinf} \\
 V_{\infty}\: &=& \:  \left(1 + \frac{2  H_0}{mc^2}\right)^{-1/4},
\label{vinf} \\
{\hat p}_K (x;\,S;\,\infty) \mid_{(K\tau)=0} &=& \,-\frac{i\h}{2}\,
 V_\infty^{-1}\cdot (KD)\cdot V_\infty \, + \,
\frac{i\h}{2}\,V_\infty \cdot (KD)^{\dagger}\cdot V_\infty^{-1} \nonumber\\
 &-&
\frac{\h\zeta}{2mc^2}\, V_\infty\cdot (\tau\nabla)(\nabla K)\cdot V_\infty
\label{pinf}\\
c\ \hat p_{\tau/c}(x;\,S;\,\infty)
&=& mc^2 \left(1 + \frac{2H_0}{mc^2}\right)^{1/2}, \
\label{einf}\\
\hat q^{(i)} (x,\,S; \infty)\ &=& q_S^{(i)} (x)\ + \ \frac{1}{2}
\left[ [ V_{\infty},\ q_S^{(i)} (x)],\ V_\infty^{-1}\right].   \label{qinf}
\end{eqnarray}

It should be emphasized that these formulae are exact relativistic
ones in the sense that they are not asymptotic and valid for any value
of $c^{-1}$. However, one should keep in mind that they are take place
in the particular frame of reference determined by the symmetry
of the globally static \ri.

One can come from Eqs.(\ref{vinf}) -- (\ref{qinf}) to
the following conclusions.

1) Not only the hamiltonian $\hat H$, but also  operators of spatial
momentum, $\hat p_K $,  and  of position, $\hat q^{(i)}$, are generally
non--local, except the case of $c^{-1} = 0$, i.e. in the exact
nonrelativistic limit, when they
coincide in form with Eqs.(\ref{kl}) and (\ref{qq}).

2)  The operators of spatial projections of momentum become local
 and coincide  in form with that of geometric quantization,
Eq.(\ref{kl}), if
\begin{equation}
  [KD, H_0] = 0,    \label{kkk}
\end{equation}
what means  that $ K^\alpha (x)$ is a Killing vector of each
level hypersurface $S$ with respect to metric induced by \ri  and,
as a consequence, of \ri itself.

3) The operators $\hat q^{(i)}$  of position on  $S$ are
noncommutative, again  except the case    $c^{-1} = 0$
or when functions $q^{(i)} (x) $ are a Cartesian coordinates $ x^i $
on a space--like hyperplane in \eu.

In the case of the globally static \ri   the  distinction between
the field--theoretically determined  operators of observables and
those that are postulated in  immediate quantization
of mechanics may not be related to the processes of particle creation
and annihilation by the external field. \\

Now consider the simplest case of the inertial frame of reference
in \eu, what means that $\Sg $  is a hyperplane $E_3$.
It is easy to see that in  this case  \eu the space
$\Phi^- (E_3;\, \infty) $ determined by $\Psi $
is a linear envelope of the negative--frequency exponentials.
Then, if  $q^{(i)} \stc x^i,\quad x^i $ being Cartesian coordinates
on $E_3 $ and $K^\alpha_{(i)} = \delta^\alpha_i $, it follows from
from Eqs.(\ref{pinf}) and (\ref{qinf}) that
\begin{equation}
\hat q^{(i)} (x;\, E_3;\, \infty) \equiv \hat x^i = x^i \,{\bf 1},
\qquad \hat p_i = \,  - i \h\frac{\ptl}{\ptl x^i},   \label{x}
\end{equation}
i.e. the canonical expressions.

However, the relativistic corrections to the position operator are nonzero
even in \eu and th inertial frames of reference,
if curvilinear  coordinates on $ E_3$ are taken  as
$q^{(i)}(x\,;E_3;\,\infty)$.
Let, e.g., $\ q^{(1)}\, =\, r $, the ordinary radial coordinate.
Then, for $ <r> \gg \hbar/mc $, one has from Eq.(\ref{qinf})
\begin{equation}
\hat r = r + \left(\frac{\hbar}{2mc}\right)^4\
\frac{1}{r^3}\,\triangle_{S_2}
+ O \left(\left(\frac{\hbar}{2mcr}\right)^6 \right), \label{rr}
\end{equation}
where $\triangle_{S_2}$ is the laplacian  on the sphere.
Apparently,  Eqs.(\ref{x}), (\ref{rr})  mean
also  that  radial positions of a quantum particle
determined  by a direct measurement of the coordinate $r $  and
calculated after measuring of three Cartesian coordinates
$x^i$ will differ if the relativistic corrections are taken into
account. Speculatively the matter looks as if the  measurement of the
distance between a spherical radar and a quantum particle would give a
result which is different from the result of locating the particle
by a huge three--dimensional wire chamber and subsequent calculation
of the distance.

The operator $\hat x^i$,  Eq.(\ref{x}), can be transformed to  the
Newton--Wigner operator [25] for to the following reason which is valid
for the general \ri.  It is easy to see that
$\varphi(x)\in \Phi^- (\Sg;\ N)$
corresponding to $\psi(x)\in \Psi (\Sg;\ N) $  satisfies the equation:
\begin{equation}
i\h\ T \ \varphi(x) \ = \ (mc^2 + H_N)\ \varphi(x),  \label{tt}
\end{equation}
so that owing to Eq.(\ref{vv})
\begin{equation}
 \{\varphi_1,\ \varphi_2 \}_S\ = \ <\varphi_1,\ \varphi_2 >_S\ \stc \
 \frac{2mc}{\h} \int_S d\sigma\
\ov\varphi_1 \left(V_N \cdot V_N^\dagger\right)^{-1}\ \varphi_2 ,
\end{equation}
The operators  of position $ \check q^{(i)}$  with respect to
the   new scalar product $<.\,, .>$
in $ \Phi^- (\Sg;\ N)$ can be introduced  by the relation
\begin{equation}
(\psi_1,\ \hat q^{(i)} \psi_2)_S
\equiv \{\varphi_1,\ q^{(i)}  \varphi_2 \}_S
\stc <\varphi_1,\  \check q^{(i)}\varphi_2 >_S
\end{equation}
Hence  it follows that
\begin{equation}
\check q^{(i)} =  q^{(i)} + \frac{1}{2mc^2} V_N^2 \cdot
[ q^{(i)},\, H_N] + \Oc.
 \label{chq}
\end{equation}
In the case when  $\rif \sim R_{1,3}$ and  $ \Sg \sim E_3 $  and
$q^{(i)} (x) \equiv x^i ,\ x^i $ being  Cartesian coordinates on
$ E_3 $, Eq.(\ref{chq})    reads  as
\begin{equation}
\check  x^i =  x^i + \frac{\ \hbar}{m (mc^2 + H_0)}
\ \frac{\partial}{\partial x^a},
\end{equation}
and $   x^i$ thus defined is just the  Newton--Wigner operator [13]  of
position in the $ x $--representation.  Therefore one may consider
Eq.(\ref{chq}) as a generalization of the  Newton--Wigner operator to \ri.

However,  introduction of  the Riemannian background becomes
natural for covariant consideration  of curvilinear coordinates  and curved
initial hypersurfaces $\Sg$ even in \eu.\\

\noindent
{\large\bf 7. Note on the Hegerfeldt Theorem}\\

Of course, the representation space $\Psi $ is in an one-to-one
correspondence  with the space $\Phi^-$  of solutions of
Eq.(\ref{r}) spanned by the negative--frequency exponentials in the sense
that any $\varphi (x) \in  \Phi^-$  can be represented in
Cartesian coordinates in the form of Eq.(\ref{az})
\begin{equation}
\varphi (x) \
= \ \sqrt{\h/2mc }\ exp\left(-i \frac{mc}{\h} x^0 )\right) \
 V_\infty   \psi (x).
\end{equation}
However, the correspondence is obviously nonlocal owing to the operator
$V_\infty$. This nonlocality is apparently a manifestation
of a paradox in quantum theory  which is referred sometimes to as
the Hegerfeldt theorem  and, in application to a single
particle,  consists in that  its wave function  having initially
a compact support  $X \subset E_3 $,  acquires nonzero values
at space--like intervals
from  $X$ in  subsequent  moments of time. This looks  as a
nonzero probability of superluminal propagation of  particles.
Hegerfeldt and  Ruijsenaars [25]  proposed a resolution of the paradox
consisting in that a localization  in a compact domain  is not possible at
all,  but, in application to our case,  they meant the localization
in terms of the field $\varphi (x)$.   However, the probability density
of localization of a particle is determined by
the field $\psi (x) $.  The initial data for  $\varphi(x)$
are related to the initial $\psi (x) $ nonlocally by
Eqs.(\ref{hinf}), (\ref{v}), (\ref{tt}). Therefore, even
if  the particle is  localized in the quasinonrelativistic sense
which is apparently the  unique  correct sense,
nevertheless, the corresponding initial data for the  relativistic field
are smeared out over  the whole $E_3$.\\

{\large\bf 8. Concluding Remarks}\\

An essential feature  of the approach exposed here  is
that the $L_2 (\Sg;\, C)$ structure of  the representation space and
operators of observables acting on it  are traced to the corresponding
field--theoretical notions of a  number of quanta, the energy--momentum and
"the position of a quantum". The latter which  leads  to the most
inconvenient conclusions on  noncommutativity of operators of coordinates
 may seem  rather  artificial but it is unique and necessary
for an intrinsic  congruence of the approach.  Indeed,
why could not one choose such sets from the limits
of these  operators for $c^{-1} = 0 \ (N = 0 )$, i.e.
simply as multiplications by functions $q^{(i)}$, which just
suggest by the canonical and geometrical quantization?  Of course,
one could, but in  QFT no physically sensible quantity would correspond
to these  operators. Particularly,  one should then  refuse the convenient
definition of  conserved quantities by Eq.(\ref{pk}) following
from the  Noether theorems and equivalence between the energy
operator $\hat p_{\tau/c}  $ and  the hamiltonian $\hat H$.

The present approach and that of quantization of
mechanics   are together obviously  a manifestation of the wave--corpuscular
dualism in quantum theory. The former corresponds to the point of view
that the Schr\"odinger wave function $\psi (x)$ is not only a mathematical
object but related to the  field $\varphi (x)$ carrying an energy--momentum.
At the same time, it leads to the inconvenient  conclusion that
in an  external field  generally  neither operators of momentum nor of
coordinates generate a complete set of commuting observables.

Another point for doubts on the presented scheme might be the question of
completeness of  the spaces $\Phi^- (\Sg;\, N) $ and $\Psi (\Sg;\, N) $
since $\varphi (x)$ and $\psi (x)$ are subordinated to conditions of
validity of the  asymptotic expansions. Of course, this question
needs an investigation as well as many other points where the
adjective "asymptotic" is used (the asymptotic inner product, asymptotic
Hermiticity, asymptotic unitarity,  the range of validity  the asymptotic
expansions  along  the frame of reference  etc.).  However, the situation
looks  not worse than with the standard NRQM
which,  in fact,  is also a limit of a more general relativistic  theory,
but nevertheless its mathematically refinement is developed as
it were a closed self--consistent theory. Besides, in the case of the
globally static \ri our construction is not asymptotic and in this sense
it is closed.

There are questions which are more specific to  quantum mechanics
in \ri. For example,  generally  the frame of reference $\tau^\alpha (x)$
has focal points in the future or past  (or both) of $\Sg$, in which the
normal geodesics  from different points of  $\Sg $ intersect and
the solution of the Cauchy problem  for the Hamilton--Jacobi equation
 (\ref{hj}) has a singularity. The question arises: Can a  method
of extension of an asymptotic solution over the point be
elaborated for the quasinonrelativistic asymptotics  analogous to that by
Maslov and Fedoriuk [22], for the quasiclassical one? A simple
instance when this problem should be studied  is  quantum
mechanics in \eu  determined by a curved $\Sg$  instead of
a convenient $E_3$.

Another interesting direction of study  is  to consider nongeodesic normal
frames of reference defined by a  Hamilton--Jacobi equation
different from Eq.(\ref{hj}) thus distributing
an action of external nongravitational forces between the frame of
reference and the quantum dynamics of a particle in it.

An important  question is: what relation  exists  between   quantum
mechanics' determined by  different Cauchy hypersurfaces
$\Sg$? The operators of observables  are determined  only in
each $ \Psi^- (\Sg;\, N) $  and  can be transformed to the corresponding
$\Phi^- (\Sg;\, N) $. Each quasinonrelativistic quantum mechanics
thus defined forms a coherent, or irreducible  lattice, see [23], Sec.8-2,
and the quantum   principle of superposition takes place in it.
 On the other hand, a closure of the set theoretical union
$ \bigcup \Phi^- (\Sg_n;\, N)$ for two or even
infinite number of  essentially  different hypersurfaces $\Sg_n$ has also
the structure of vector space since any superposition of wave functions
from $ \bigcup \Phi^- $ is again a solution of the field equation (\ref{r}).
However, if one takes $\varphi_1$ and $\varphi_2$ from different
spaces $\Phi^- (\Sg_n;\, N)$, then there one may expect (and draft
calculations support this expectation though a rigorous proof
do not seem easy for the general case)  the sesquilinear functionals
$\{\varphi_1,\, \varphi_2 \}_\Sg$,
$P_K (\varphi_1,\, \varphi_2;\, \Sg)$ and
$\{\varphi_1,\, q_\Sg^{(i)} \varphi_2 \}_\Sg$
defined by Eqs.(\ref{N1}), (\ref{pk2}),
and (\ref{Q1}) asymptotically vanish. If it is the case,
then a superposition of $\varphi_1$ and $\varphi_2$ is corresponds to
asymptotically mixed state.
The situation is just  as if the spaces $\Phi^- (\Sg_n;\, N) $  form {\it
superselection sectors} in $ \bigcup \Phi^- (\Sg_n;\, N)$ which is
a reducible lattice. If this understanding is correct,  one reveals a very
interesting class of superselection rules associated to frames of reference.

 Having recalled  the quantum field--theoretical origin of the presented
construction, one could  also attempt to connect the sets of operators
of creation and annihilation of particles, which are determined on
two different Fock spaces
$ {\cal F}_a,\ a = 1,\, 2,\ $  corresponding to  given spaces
$\Phi^- (\Sg_a ;\, N) $, by a Bogoliubov transformation. Apparently,
this is possible for  sufficiently smooth metrics of \ri, and  then
a one--particle state, say, in  ${\cal F}_1$  will be
represented by a superposition of an infinite set of
different many--particles states  in ${\cal F}_2$.

At last, it is interesting to apply the field--theoretical approach
exposed here to fields of nonzero spin.
Since the operators of spin projections are defined by the lagrangian of
the field  one may expect that their algebraic properties
are different from the standard ones if $N>0$.

These problems may be criticized as academical ones. However, I think
that without study of them our knowledge of  quantum
theory would be essentially incomplete. \\

\noindent
{\bf Acknowledgement}\\
The author is  grateful to Prof. B.M.Barbashov and  Prof. I.T.Todorov
for useful discussions and to Prof. A.V.Aminova and Dr. D.V.Kalinin for
explanations of the geometric quantization. \\

{\large\bf Appendix} \\

The proof of unitary equivalence of the operator energy, Eq.(\ref{E}),
to the hamiltonian  in the \Sche (\ref{t2})  is
the same for any value of $\zeta$. Therefore
consider for brevity  the case of $\zeta = 0$.

  Using expressions  Eq.(\ref{ham}) for $\hat H_N$\, Eq.(\ref{E})
and relation Eq.(\ref{vv} for $V_N $, one can rewrite  Eq.(\ref{u})
as an equation for $V_N$:
\begin{eqnarray}
 [i\h T, V_N]  =
 \left\{ H_N - \left(1 + \frac{H_N^\dagger + H_N }{2 mc^2}\right)^{-1}
 \left(H_0\
+\ \frac{\tilde H_N^\dagger\cdot \tilde H_N }{2 mc^2}\right)\right\}
 \cdot  V_N \label{4t}  \ +\ \Oc. \label{tv}
\end{eqnarray}
Since the operator $V_N$ is a polynomial of the space derivatives
 $D_\alpha$ with  coefficients
depending on  $ x $\,  Eq.(\ref{4t}) is equivalent to a
linear evolution system along the field $\tau^\alpha$\
on these coefficients  in virtue of the relation
\begin{equation}
 [T,\ D_\alpha] \ =\  \tau^\gamma [\nabla_\gamma,\ \nabla_\alpha]\ + \
\nabla_\alpha \tau^\gamma D_\gamma\
+ \ \frac{1}{2} D_\alpha \nabla_\gamma \tau^\gamma
\end{equation}
The first term at the right--hand side  will    generate in Eq.(\ref{4t})
a term proportional to the Riemann--Christoffel curvature tensor and
consequently the commutator  is again a polynomial of  $D_\alpha $ of
the same order.  A solution of the Cauchy problem for this evolution system
always exists in some neighborhood of an initial $ S = \Sg\ $. However,
the solution should satisfy  to the condition  (\ref{vv}). The latter,
 having been imposed  the Cauchy data on $\Sg$, is fulfilled
$\tau^\alpha$ because
\begin{equation}
\biggl[\tau^\alpha \ptl_\alpha,\
V_N^\dagger\cdot \left(1
+ \frac{H_N^\dagger + H_N }{2 mc^2}\right)\cdot V_N \biggr]\ =\ \Oc
\label{4v}
\end{equation}
in virtue of Eq.(\ref{4t}) and the condition itself. Performing the
commutations, one may easily verify  that Eq.(\ref{4v}) is equivalent to
the condition of the asymptotical  hermiticity of the  hamiltonian
$\hat H_N$ expressed in terms of  $H_N$.\\

 {\large\bf References}\\

\noindent
1.\ Feshbach H., Villars F. (1958), {\it Rev.Mod.Phys.},{\bf 30},  24.\\
2.\ Gibbons G.W., Pohle H.J. (1993), {\it Nucl.Phys.}, {\bf B410},  117.\\
3.\ Hobbs J.M., (1968), Ann.Phys. {\bf 47}, 141.\\
4.\ DeWitt B.S.,  Brehme R.W. Ann.Phys. (1960), {\bf 9}, 220.\\
5.\ Schweber S. (1961), {\it An Introduction to Relativistic Quantum Field
Theory}, Row, Peterson and Co, N.Y.\\
6.\ Gorbatzevich A.K. (1985), {\it Quantum Mechanics in General Relativity}
(in Russian), Belorussian State University Press, Minsk. \\
7.\  Stephani H. (1965), {\it Ann.d.Physik}, Bd.15, S. 12.\\
8.\ Tagirov E.A. (1990), {\it Theoretical and Mathematical Physics},
{\bf 84}, N 3, 419--430 .\\
9.\ Tagirov E.A. (1992), {\it Theoretical and Mathematical Physics},
{\bf 90}, N 3, 412--423 .\\
10. Tagirov E.A. (1996), {\it Theoretical and Mathematical Physics},
{\bf 106}, N 1, 99 -- 107.\\
11. \'Sniatycki J., (1980), {\it Geometric Quantization and
Quantum Mechanics}, Springer--Verlag, New York, Heidelberg,  Berlin.\\
12. Dirac P.A.M. (1958), {\it The  Principles of Quantum Mechanics},
4th ed., Clarendon Press, Oxford.\\
13. Chernikov N.A., Tagirov E.A. (1969),  {\it Annales de l'Institute
Henry Poincar\`e}, {\bf A9},  39.\\
14. Tagirov E.A. (1973), {\it Annals of  Physics (N.Y.)} {\bf 76 }, 561. \\
15. Grib A.A.,  Mamaev S.G., Mostepanenko V.M. (1980), {\it Quantum Effects
in Intensive External Fields} (in Russian), Atomizdat, Moskow.\\
16. Birrell N.D., Davies P.G.W. (1982), {\it Quantum Fields in Curved Space},
Cambridge  University  Press , Cambridge. \\
17. Polubarinov I.V. (1973),  {\it JINR Communication}, P2--8371, Dubna.\\
18. Weinberg S. (1989), {\it Annals of  Physics (N.Y.)} {\bf 194 }, 336. \\
19. Bronnikov K.A.,  Tagirov E.A. (1968), {\it JINR Communication}
P2-4151, Dubna.\\
20. Parker L. (1969) {\it Physical Review}, {\bf 183}, 1057.\\
21. Grib A.A.   Mamaev S.G. (1969), {\it Journal of Soviet Nuclear
Physics,} {\bf 10}, 1276.\\
22. Maslov V.P.  Fedoriuk M.V. (1976), {\it Quasiclassical Approximation for
Equations of Quantum Mechanics} (in Russian), Nauka, Moscow. \\
23. Jauch J.M. (1968), {\it Foundations of Quantum Mechanics},
Addison--Wesley, Reading, Massachusetts.   \\
24. Blokhintsev D.I. (1966),   {\it JINR Communication}, P2--2631, Dubna.\\
25. Newton T.D., Wigner E.P. (1949), {\it Rev.Mod.Phys.} {\bf 21}, 400.\\
26. Hegerfeldt G.H.,  Ruijsenaars S.N., (1980)  {\it Phys.Rev. D }
{\bf 22}, 377. \\
\end{document}